\documentclass{jfm}

\usepackage{graphicx}
\usepackage{amsmath, amsfonts, amsxtra, amssymb, amsbsy}
\usepackage{overpic}
\usepackage{bm,color}
\usepackage[sort]{natbib}
\usepackage[dvipsnames]{xcolor}

\usepackage{tikz}

\usepackage{my_macros}

\numberwithin{equation}{section}

\graphicspath{{./}{figures/}}

\title{A multiscale method to calculate filter blockage}
\author[M.~P.~Dalwadi, M.~Bruna, and I.~M.~Griffiths]{M.~P.~Dalwadi$^{1,2, a}$, M.~Bruna$^{2, b}$, and I.~M.~Griffiths$^{2, c}$}
\affiliation{\small $^{1}$ Synthetic Biology Research Centre, The University of Nottingham, University Park, Nottingham, NG7 2RD, \\
\small $^{2}$ Mathematical Institute, University of Oxford, Oxford, OX2 6GG, \\ 
\small $^{a}$ email: mohit.dalwadi@nottingham.ac.uk 
\small $^{b}$ email: bruna@maths.ox.ac.uk 
\small $^{c}$ email: ian.griffiths@maths.ox.ac.uk}

\begin{document}

\maketitle

\begin{abstract}
Filters that act by adsorbing contaminant onto their pore walls will experience a decrease in porosity over time, and may eventually block. As adsorption will generally be larger towards the entrance of a filter, where the concentration of contaminant particles is higher, these effects can also result in a spatially varying porosity. We investigate this dynamic process using an extension of homogenization theory that accounts for a macroscale variation in microstructure. We formulate and homogenize the coupled problems of flow through a filter with a near-periodic time-dependent microstructure, solute transport due to advection, diffusion, and filter adsorption, and filter structure evolution due to the adsorption of contaminant. We use the homogenized equations to investigate how the contaminant removal and filter lifespan depend on the initial porosity distribution for a unidirectional flow. We confirm a conjecture made in \citet{dalwadi2015understanding} that filters with an initially negative porosity gradient have a longer lifespan and remove more contaminant than filters with an initially constant porosity, or worse, an initially positive porosity gradient. Additionally, we determine which initial porosity distributions result in a filter that will block everywhere at once by exploiting an asymptotic reduction of the homogenized equations. We show that these filters remove more contaminant than other filters with the same initial average porosity, but that filters which block everywhere at once are limited by how large their initial average porosity can be.
\end{abstract}

\section{Introduction}
  
Filtration is a vital process in many industries, such as kidney dialysis \citep{lonsdale1982growth}, air purification \citep{barhate2007nanofibrous}, waste water treatment \citep{vandevivere1998review}, and beer production \citep{fillaudeau2002yeast}. Although the industrial applications may vary widely, the main goal is often the same: to maximize the removal of contaminants or particulates entrained within the fluid that passes through the filter. Since a single experiment may take hours to complete, mathematical modelling provides a valuable tool for assisting with filter design.

Broadly speaking, filters are composed of either material containing a series of pores (e.g., track-etched membranes) or a fibrous mesh (e.g., so-called depth filters). In the former case, if the contaminants are larger than the pores, filtration occurs at the pore entrance via size exclusion. A \emph{cake layer} of particulates builds up on the entrance, making it more difficult for fluid to pass through the filter. When the contaminants are smaller than the pores, they are able to penetrate into the pores, where they may become trapped in the tortuous pore network and adhere to the pore walls. This reduces the available space through which contaminated fluid can flow, and again makes it more difficult for fluid to pass through the filter~\citep{griffiths2014combined}. Filters composed of a fibrous mesh provide an internal structure of obstacles to which the contaminants can adsorb. As with filters composed of pores, the physical trapping of particulates within the filter will cause a reduction in the available space through which the fluid can flow. Since the contaminants are removed as the fluid flows into the filter depth, their concentration decreases with depth and so this constriction effect is usually more significant towards the filter entrance~\citep{datta1998gradient}. Over time, this constriction may ultimately become a blockage, halting filtration. To prolong lifespan, filters are often designed such that their porosity decreases with depth, allowing areas of the filter away from the entrance to have trapped more contaminants by the time of blocking~ \citep{burggraaf1991synthesis,barg2009processing,anderson1951filter,dickerson2005gradient,vida2012characterization}. These are known as \emph{porosity-graded filters}. Moreover, an initially homogenous filter will become graded over time as the local porosity changes due to local contaminant trapping.

While mathematical methods can provide a cost-effective way to predict solute transport without experimentally sweeping through parameter space, the explicit coupling of solute transport with filter evolution yields a complicated moving boundary problem. This significantly increases the mathematical and computational effort required to solve the system and, as such, many mathematical models of filtration consider only one of these mechanisms. That is, either the problem of solute transport, or the problem of filter evolution. For example, in \citet{griffiths2014combined}, the filter evolution is considered by simulating a \emph{microscale} model (on the lengthscale of pore size) for the blocking of a network of pores from a fundamental mechanistic level. While the results are consistent with the standard set of constitutive \emph{macroscale} laws (on the lengthscale of filter size) used to predict fluid throughput (often fitted with experimental data)~\citep{bowen1995steps}, the flow problem is simplified by assuming Poiseuille flow through the constricting cylindrical pores. This idea is generalized to explore the effect of tapered pores and filters composed of multiple membrane layers in \citet{griffiths2016designing}. A different approach to the problem of solute transport past sinks is used in \citet{chernyavsky2010mathematical}, where the authors considered blood flow and nutrient delivery in the placenta. In this paper, the transport of nutrient governed by diffusion and advection with unidirectional flow past randomly placed point sinks in the placenta is considered in one dimension. The effect of placental growth and the nature of the local flow are neglected to focus on the important features of the paper, such as quantifying the error in upscaling techniques. The authors do this by analytically determining the concentration field for different statistical distributions of the sinks using homogenization techniques and comparing these to exact solutions.

Mathematical homogenization provides an upscaling method whereby the fundamental behaviour of concern, such as the contaminant removal rate, is captured without the computational expense of globally calculating the microscale behaviour. This is achieved by averaging the microscale variation while retaining the macroscale variation. For this procedure to be valid, it is necessary for the ratio between microscale and macroscale lengthscales to be small. Although traditional homogenization techniques also require the microstructure to be strictly periodic, it can be extended to microstructures that are only \emph{near-to} periodic \citep{richardson2011derivation}. However, the added generality of this extension comes with a drawback. A varying microstructure means that the cell problem must be solved at every point in the macroscale, increasing the computational expense required to solve the problem. This issue is bypassed in~\citet{Bruna2015diffusion}, by considering a microstructure consisting of an array of spherical obstacles whose radii vary spatially over the macroscale length. Imposing a specific one-parameter shape on the microstructure means that the cell problem is identified by a single parameter -- the cell porosity, and the macroscale equation can be written explicitly as a function of the porosity.

In a previous work \citep{dalwadi2015understanding}, we considered solute transport through a porosity-graded filter. The filter was modelled as a series of fixed spherical obstacles with a near-to-periodic spatial variation on the macroscale, through which the contaminated fluid flowed. The techniques developed in \citet{Bruna2015diffusion} were used to derive a homogenized advection--diffusion macroscale equation. Although the particles that make up the solute would physically accumulate on the surface of each obstacle, causing the filter geometry to vary over time, we assumed that the particles were sufficiently small and dilute that this effect was negligible on the timescale of interest. Thus, we did not consider the filter evolution or the subsequent filter blocking. We showed that filters whose porosity decreases with depth have a much more uniform adsorption than those with increasing porosity. Moreover, we conjectured that such uniform adsorption would prolong filter lifespan, as contaminant removal would be distributed throughout the filter rather than near the filter entrance, and thus blocking may occur `all at once' rather than in just one place. The aim of this paper is to confirm this conjecture, to quantify the time-dependent evolution of the filter, and to determine how to construct a filter that will block all at once.

In this paper, we derive a generalized mathematical model for a porosity-graded filter which explicitly accounts for changes in the filter microstructure over time due to particle adsorption. The time evolution of the filter geometry means that the full fluid and transport problems are coupled. We use the homogenization method developed in \citet{richardson2011derivation}, \citet{Bruna2015diffusion}, and \citet{dalwadi2015understanding} to account for a microstructure that varies over the macroscale and to systematically determine an effective macroscale equation. The filter material is modelled as a lattice of obstacles, whose size varies both spatially over a long lengthscale and temporally, past which a fluid with suspended contaminant particles flows. The particles are transported via advection and diffusion, and can be trapped on the filter as they come into contact with the obstacles. The accumulation of contaminant particles via trapping modifies the filter geometry, and this process eventually leads to pore blockage within the filter. We assume that the contaminant particles are small and in dilute suspension within the fluid. Thus, particle--particle interactions are negligible, and the dominant interaction is the adsorption of contaminant on the obstacles, which we assume is proportional to the concentration of contaminant at the surface.

We present the full flow and contaminant transport problems in \S\ref{sec: Model desc}, and homogenize these to obtain effective equations on the macroscale in \S\ref{sec: Homog}. In \S\ref{sec: Uni-dir problem} we investigate the effect of blocking by considering a filter whose porosity varies in one direction only, with a unidirectional flow in the same direction as the porosity gradient. We use asymptotic methods to significantly reduce the computational expense of solving our system, allowing us to perform an efficient parameter sweep, and we use our results to validate the conjectures made in \cite{dalwadi2015understanding}. Additionally, these asymptotic results allow us to solve the inverse problem of determining which initial porosity distributions result in filters that block everywhere at once. We conclude in \S\ref{sec: Discussion} with an overview and discussion of our results, and we discuss the challenge of determining the filter that globally optimizes contaminant removal.

\section{Model description}
\label{sec: Model desc}

We consider the transport of contaminant particles via advection and diffusion through a porosity-graded filter, and its resulting blocking. The filter is modelled as a collection of infinitely long solid cylindrical fibres with circular cross-section, whose axes are parallel and where the fluid flows normal to these axes. Contaminants can adsorb onto the solid fibre surface (obstacle surface) and, over time, the build-up of contaminant will result in a reduction in porosity. We describe the contaminant particle distribution in terms of the concentration $\ndc(\ndx,\ndt)$, where $\ndc$ is the number of moles of contaminant per unit volume, $\ndx$ is the spatial vector coordinate, and $\ndt$ is time. 

Although the problem we have described is three-dimensional, the inherent symmetry allows us to consider a two-dimensional problem as follows. The concentration field is defined within the fluid phase of the domain $\Regf(\ndt) \subset \mathbb{R}^2$. We denote $\Rege \subset \mathbb{R}^2$ as the entire domain, which we refer to as the \emph{porous medium}. We define the solid phase of the domain as $\Regs \subset \mathbb{R}^2$, noting that $\Regs(\ndt) = \Rege \setminus \Regf(\ndt)$, and we define the boundary between fluid and solid phases, $\partial \Regf(\ndt)$, which we refer to as the obstacle surface.

\begin{figure}
\centering
    	\includegraphics[width=\textwidth]{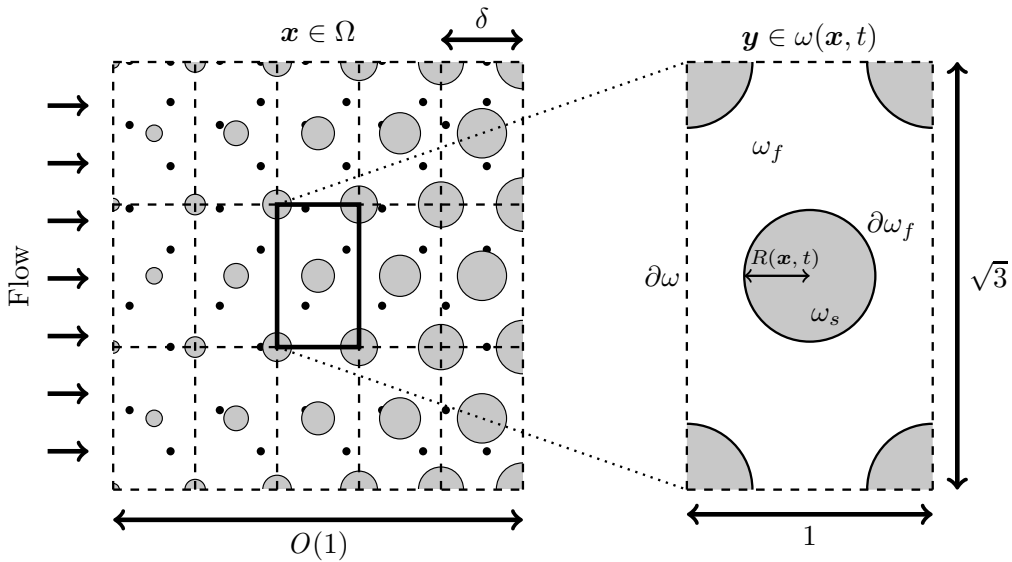}
\caption{Model schematic in two dimensions. Left: An example in which the macroscale porosity decreases in the direction of the flow. Right: A magnified view of a given cell $\rege(\bs{x},t)$, with microscale coordinate $\bs{y} \in \left[-1/2,1/2\right] \times \left[-\sqrt{3}/2,\sqrt{3}/2 \right]$.}
\label{fig: Set-up schematic}
\end{figure}

The solid phase is modelled by a collection of fixed non-overlapping disks, whose centres are located on a hexagonal lattice at a distance $\epsa \leng$ apart, where $\epsa$ is a (small) dimensionless parameter and $\leng$ is the characteristic filter depth. We allow the radii of the circles to vary in both space and time, and a circle with centre at $\ndx$ has radius $\epsa \tilde{\rad}(\ndt;\ndx)$, where $2 \tilde{\rad} \leq \leng$ and blocking occurs when equality is reached. Thus, blocking does not correspond to the fraction of solid phase reaching $1$. We use a hexagonal lattice in contrast to the cubic lattice used in \cite{dalwadi2015understanding} so that we can reach a higher solid fraction at the point of blocking: arranging the obstacles on a cubic lattice, we can only reach a blocking fraction of $\pi/4 \approx 0.785$, whereas the hexagonal lattice at contact can reach a blocking fraction of $\pi /(2\sqrt{3}) = 0.907$. A schematic of this set-up is shown in the left-hand side of figure~\ref{fig: Set-up schematic}.

The pore space is assumed to be entirely saturated by an incompressible Newtonian fluid, which satisfies the Stokes equations,
\begin{subequations}
\label{eq: Flow problem}
\begin{alignat}{2}
-\ndn \ndp + \viscosity \ndn^2 \ndu &= \zv,& \qquad  &\ndx \in \Regf(\ndt), \\
\ndn \bcdot \ndu &= 0,& \qquad  &\ndx \in \Regf(\ndt), \\
\label{eq: Flow problem BC dim}
\ndu &= -\epsa \frac{\partial \tilde{\rad}}{\partial \ndt} \bs{n},& \qquad &\ndx \in \partial \Regf(\ndt),
\end{alignat}
\end{subequations}
where $\ndn$ refers to the nabla operator with respect to $\ndx$, $\viscosity$ is the constant fluid viscosity, $\ndu(\ndx,\ndt)$ is the fluid velocity, $\ndp(\ndx,\ndt)$ is the fluid pressure, and $\bs{n}$ is the unit normal to the obstacle surface directed into the obstacle.

We assume that the contaminant diffuses within the fluid with a constant diffusion coefficient, $D$, and is advected by the velocity field. The governing equation is thus
\begin{align}
\label{eq: transport problem}
\pbyp{\ndc}{\ndt} = \ndn \bcdot \left( D \ndn \conc - \ndu \ndc \right), \qquad  \ndx \in \Regf(\ndt).
\end{align}

To determine the correct boundary condition for the concentration on a moving boundary, it is helpful to consider the rate of change of the total number of moles of contaminant in an arbitrary time-dependent volume $V(t)$ with boundary $\partial V(t)$:
\begin{align}
\label{eq: RTT}
\dbyd{}{\ndt} \int_{V(\ndt)} \! \ndc \, \mathrm{d}V &= \int_{V(\ndt)} \! \pbyp{\ndc}{\ndt} \, \mathrm{d}V + \int_{\partial V(\ndt)} \!  \left( \bs{n} \bcdot \tilde{\bs{v}} \right) \ndc \, \mathrm{d}S = \int_{\partial V(\ndt)} \!\bs{n} \bcdot \left( D \ndn \ndc + \left(\tilde{\bs{v}} - \ndu \right)\ndc\right)  \, \mathrm{d}S,
\end{align}
where $\tilde{\bs{v}}$ is the velocity of the boundary and $\bs{n}$ is the outward unit normal to $\partial V(t)$. In \eqref{eq: RTT}, we have used the divergence theorem and \eqref{eq: transport problem}. As $\ndu = \tilde{\bs{v}}$ on $\partial V(t)$, \eqref{eq: RTT} reduces to
\begin{align}
\label{eq: rate of change of conc}
\dbyd{}{\ndt} \int_{V(\ndt)} \! \ndc \, \mathrm{d}V = \int_{\partial V(\ndt)} \!  \bs{n} \bcdot D \ndn \ndc \, \mathrm{d}S.
\end{align}

We suppose that the rate of change of the concentration is linearly dependent on the concentration at the obstacle surface. Thus, using \eqref{eq: rate of change of conc} in an infinitesimal volume containing a fragment of the obstacle surface, we obtain a partially adsorbing Robin boundary condition:
\begin{align}
\label{eq: dim conc BC}
- \kn \ndc  &= \bs{n}\bcdot  D \ndn \ndc, \qquad \ndx \in \partial \Regf(\ndt),
\end{align}
where $\kn \geqslant 0$ is the constant contaminant-adsorption coefficient. There is no adsorption when $\kn = 0$, and the adsorption is instantaneous in the limit as $\kn \to \infty$. 

Finally, we must couple the growth of the obstacles to the accumulation of contaminant at the obstacle boundary. This is in the form of a volume conservation law, where the volume of contaminant lost is equal to the volume gained by the obstacle. Modifying \eqref{eq: RTT} to consider volume instead of concentration, it follows that the movement of the interface in the normal direction due to contaminant adsorption is
\begin{align}
\label{eq: Constitutive law}
\epsa \pbyp{\tilde{\rad}}{\ndt} = - \Vm \bs{n} \bcdot D \ndn \ndc, \qquad \ndx \in \partial \Regf(\ndt),
\end{align}
where $\Vm$ is the molar volume of contaminant in the filtrate. Here, we have assumed that the adhesion of contaminant causes, on average, a change in the radius of the obstacles. We have included the effect of filter porosity changing due to contaminant particle adsorption but neglected the effect of particle--particle interactions because the contaminant particles are small compared with the obstacles and in dilute suspension within the fluid, hence particle--particle interactions occur far less frequently than particle--obstacle interactions. The interfacial condition \eqref{eq: Constitutive law} is similar to a Stefan condition for phase-change problems in heat transport \citep{gupta2003classical}.

\subsection{Dimensionless equations}
\label{sec: Dimensionless equations}

We scale the variables via $\ndx = \leng \bs{x}$, $\ndu = \velscale \uu$, $\ndt = ( \epsa \leng / ( c_{\infty}\Vm \kn))t$, $\tilde{\rad} = \leng \rad$, $\ndc = c_{\infty} \conc$, and \linebreak $\ndp = (\viscosity \velscale/(\epsa^2 \leng))p$, where $\velscale$  and $c_\infty$ are characteristic velocity and concentration scales respectively. The time scale is chosen to balance obstacle growth with contaminant concentration in \eqref{eq: Constitutive law}, and the pressure scale is chosen to balance the pressure gradient over the macroscale with viscous forces over the obstacle microscale. The flow problem~\eqref{eq: Flow problem} then transforms to 
\begin{subequations}
\label{eq: Flow problem ND}
\begin{alignat}{2}
-\nabla p + \epsa^2\nabla^2 \uu &= \zv,& \qquad  &\bs{x} \in \Regf(t), \\
\nabla \bcdot \uu &= 0,& \qquad  &\bs{x} \in \Regf(t), \\
\label{eq: Flow BC ND}
\uu &=  -\epsa \alpha \pbyp{\rad}{t} \bs{n},& \qquad &\bs{x} \in \partial \Regf(t),
\end{alignat}
\end{subequations}
where $\alpha = c_{\infty} \Vm \kn/(\epsa \velscale)$.

For the concentration problem given by \eqref{eq: transport problem}, \eqref{eq: dim conc BC}, and \eqref{eq: Constitutive law}, we obtain the dimensionless solute-transport equation
\begin{subequations}
\label{eq: transport problem ND}
\begin{alignat}{2}
\alpha \pbyp{\conc}{t} &= \nabla \bcdot \left( \Pec^{-1}\nabla \conc - \uu  \conc \right),& \qquad  &\bs{x} \in \Regf(t) , \\
\label{eq: transport problem ND BC}
-\epsa \kk \conc &= \bs{n}\bcdot  \Pec^{-1}\nabla \conc ,& \qquad &\bs{x} \in \partial \Regf(t), \\
\label{eq: transport problem ND growing obstacle}
\pbyp{\rad}{t} &= \conc, &\qquad &\bs{x} \in \partial \Regf(t),
\end{alignat}
\end{subequations}
where $\Pec = \velscale \leng / D$ is the P\'{e}clet number, and $\kk = \kn/(\epsa \velscale)$. The boundary condition \eqref{eq: transport problem ND BC} expresses the flux across the moving boundary relative to the boundary, and as such does not contain any velocity component. This system reduces to that considered in \citet{dalwadi2015understanding} when $\rad$ is independent of $t$, \emph{i.e.} when $\Vm = 0$, which can be seen by scaling $t \sim \alpha$ and then taking $\alpha \to 0$. We assume that $\alpha$, $\Pec$ and $\kk$ are all $\order{1}$ parameters, which corresponds to the richest asymptotic limit. That is, all mechanisms contribute at leading order, from which all asymptotic sublimits may be distilled. In practice, the contaminant accumulates on the obstacles over a much longer timescale than the fluid takes to travel through the porous medium. This manifests in the smallness of the parameter $\alpha$, and we exploit this feature in~\textsection\ref{sec: Asymptotic reductions} to consider the physically relevant effect of slow contaminant accumulation.  

In dimensionless units, the obstacles now form a two-dimensional hexagonal lattice of circles whose centres are a distance of $\epsa$ apart, and a circle with centre at $\bs{x}$ has radius $\epsa \rad(\bs{x},t)$ (figure~\ref{fig: Set-up schematic}).

\section{Homogenization}
\label{sec: Homog}

The complexity of the problem geometry is reduced by homogenizing the governing equations~\eqref{eq: Flow problem ND}--\eqref{eq: transport problem ND} using the method of multiple scales. This provides effective equations on a simpler macroscale domain, which formally capture the relevant information about the microscale geometry. Following standard homogenization theory, we introduce a microscale variable $\bs{y} = \bs{x}/\epsa$ and treat $\bs{x}$ and $\bs{y}$ as independent. This adds an extra degree of freedom which is later removed by imposing that the solution is periodic in $\bs{y}$. Hence, any small variation between unit cells is captured through the macroscale variable $\bs{x}$. As shown in the right-hand side of figure~\ref{fig: Set-up schematic}, the microscale variable $\bs y$ is defined in the unit cell $\rege(\bs{x},t)$, whereas the macroscale variable $\bs x$ spans across the whole filter. The solid portion of the cell, occupied by obstacles, is denoted by $\regs({\bs x},t)$. The fluid portion of the cell is $\regf(\bs{x},t) = \rege(\bs{x},t) \backslash \regs(\bs{x},t)$. The boundary between solid and fluid portions within the cell is denoted by $\partial \regf (\bs{x},t)$, while the outer boundary of the unit cell is $\partial \rege(\bs{x},t)$ (see figure~\ref{fig: Set-up schematic}). Further, we treat each dependent variable as a function of both $\bs{x}$ and $\bs{y}$. Thus, spatial derivatives transform in the following manner
\begin{align}
\label{eq: spatial transform}
\nabla \mapsto \nabla_{\bs{x}} + \dfrac{1}{\epsa} \nabla_{\bs{y}},
\end{align}
where $\nabla_{\bs{x}}$ and $\nabla_{\bs{y}}$ refer to the nabla operator in the $\bs{x}$- and $\bs{y}$-coordinate systems respectively. The introduction of this new spatial variable also changes the normal vector $\bs{n}$, used to evaluate the boundary conditions \eqref{eq: Flow BC ND} and \eqref{eq: transport problem ND BC}, to
\begin{align}
\label{eq: normal in microscale}
\bs{n} = \dfrac{\bs{n}_{\bs{y}} + \epsa \nabla_{\bs{x}} \rad}{\|\bs{n}_{\bs{y}} + \epsa \nabla_{\bs{x}} \rad\|},
\end{align}
where $\bs{n}_{\bs{y}} = -\nabla_{\bs{y}} \|\bs{y}\| = - \bs{y}/\|\bs{y}\|$ is the geometric outward unit normal on the obstacle boundary $\partial \regf(\bs{x},t)$, and $\epsa \nabla_{\bs{x}} \rad$ accounts for the macroscale effect of varying obstacle size. The details of this transformation are described in \citet{dalwadi2015understanding}.

As our goal is to derive effective governing equations that are valid in the macroscale domain, we consider variables averaged over an entire cell $\rege(\bs{x},t)$. To this end, we define the macroscale porosity $\porosity(\bs{x},t)$ to be
\begin{align}
\label{eq: porosity}
\porosity(\bs{x},t) = \dfrac{ |\regf(\bs{x},t)| }{ |\omega ({\bs x},t)| },
\end{align}
and the volumetric average concentration $\creal$ and volumetric average fluid velocity $\UU$ (known as the \emph{Darcy velocity} in porous-media formulations \citep{kaviany2012principles}) as follows
\begin{subequations}
\label{eq: Volumetric average}
\begin{align}
\label{eq: Volumetric average concentration}
\creal(\bs{x},t) &= \dfrac{1}{|\rege (\bs{x},t)|} \int_{\rege (\bs{x},t)}  \conc(\bs{x},\bs{y},t) \, \mathrm{d}\bs{y} = \dfrac{1}{|\rege (\bs{x},t)|}\int_{\regf (\bs{x},t)}  \conc(\bs{x},\bs{y},t) \, \mathrm{d}\bs{y}, \\
\label{eq: Volumetric average velocity}
\UU(\bs{x},t) &=  \dfrac{1}{|\rege (\bs{x},t)|} \int_{\rege (\bs{x},t)}  \uu(\bs{x},\bs{y},t) \, \mathrm{d}\bs{y} =  \dfrac{1}{|\rege (\bs{x},t)|}\int_{\regf (\bs{x},t)}  \uu(\bs{x},\bs{y},t) \, \mathrm{d}\bs{y},
\end{align}
\end{subequations}
defining $\conc = 0$ and $\uu \equiv \zv $ in $\regs({\bs x},t)$.

We first consider the flow problem~\eqref{eq: Flow problem ND}, and use the results in the solute-transport problem~\eqref{eq: transport problem ND}.

\subsection{Flow problem}
\label{App: Flow problem no growth}

Under the spatial transforms~\eqref{eq: spatial transform} and~\eqref{eq: normal in microscale}, the flow equations~\eqref{eq: Flow problem ND} become
\begin{subequations}
\label{eq: Flow problem spatial transformed}
\begin{alignat}{2}
\label{eq: Flow problem spatial transformed momentum}
- \left(\epsa^{-1} \nabla_{\bs{y}} + \nabla_{\bs{x}} \right) p + \left(\nabla_{\bs{y}} + \epsa \nabla_{\bs{x}}\right)^2 \uu &= \zv,& \qquad &\bs{y} \in \regf(\bs{x},t), \\
\label{eq: Flow problem spatial transformed cont}
\left(\nabla_{\bs{y}} + \epsa \nabla_{\bs{x}} \right) \bcdot \uu &= 0,& \qquad &\bs{y} \in \regf(\bs{x},t), \\
\label{eq: Flow problem spatial transformed no slip}
\uu &=  -\epsa \alpha \pbyp{\rad}{t}\bs{n}_{\bs{y}} + \order{\epsa^2},& \qquad &\bs{y} \in \partial \regf(\bs{x},t), \\
\label{eq: Flow problem spatial transformed periodic}
\uu&, \, p \text{ periodic},& \qquad &\bs{y} \in \partial \rege(\bs{x},t),
\end{alignat}
\end{subequations}
Expanding the flow velocity and pressure in powers of $\epsa$ as usual in the multiple-scales method (see Dalwadi et al.~2015 for details), we find that the leading-order pressure, $p_0$, is independent of the microscale, \emph{i.e.}, $p_0(\bs{x},t)$.
Considering the next order in \eqref{eq: Flow problem spatial transformed} allows us to write
\begin{subequations}
\label{eq: u_0 and p_1 scalings}
\begin{align}
\uu_0 &= - \Ku(\bs{x},\bs{y},t) \nabla_{\bs{x}} p_0, \\
p_1 &= - \pres(\bs{x},\bs{y},t) \bcdot \nabla_{\bs{x}} p_0 + \overline{p}(\bs{x},t),
\end{align}
\end{subequations}
where $\overline{p}$ is an arbitrary function of the macroscale only, and the matrix function $\Ku$ and the vector function $\pres$ satisfy the so-called \emph{cell problem}:
\begin{subequations}
\label{eq: Flow problem  order 1 after scalings}
\begin{alignat}{2}
\Iu - \nabla_{\bs{y}} \pres + \nabla^2_{\bs{y}} \Ku &= \zu,& \qquad &\bs{y} \in \regf(\bs{x},t), \\
\nabla_{\bs{y}} \bcdot \Ku &= \zv,& \qquad &\bs{y} \in \regf(\bs{x},t), \\
\Ku &= \zu,& \qquad &\bs{y} \in \partial \regf(\bs{x},t), \\
\Ku&, \, \pres \text{ periodic},& \qquad &\bs{y} \in \partial \rege(\bs{x},t).
\end{alignat}
\end{subequations}
Here, $\Iu$ is the two-dimensional identity matrix, and the dependence of $\Ku$ and $\pres$ on $\bs{x}$ and $t$ is due to the dependence of the microscale boundary $\partial \regf(\bs{x},t)$ on the macroscale variable.

Integrating~(\ref{eq: u_0 and p_1 scalings}a) over $\regf(\bs{x},t)$ and using (\ref{eq: Volumetric average velocity}), we obtain the homogenized Darcy relation 
\begin{subequations}
\label{eq: Darcy velocity}
\begin{align}
\label{eq: Darcy velocity order 1}
\UU({\bs x},t) & = -\mathcal{K}(\porosity) \nabla_{\bs{x}} p
\end{align}
at leading order, where $\mathcal{K}(\porosity)$ is a scalar function that contains the necessary microscale structure information, and is defined by
\begin{align}
\label{eq: def of K}
\mathcal{K}(\porosity)\Iu &= \dfrac{1}{|\rege (\bs{x},t)|}\int_{\regf (\bs{x},t)}  \Ku \, \mathrm{d}\bs{y}.
\end{align}
\end{subequations}
We note that the integral of $\Ku$ in~\eqref{eq: def of K} is a multiple of the identity matrix due to the symmetry of the cell problem described by~\eqref{eq: Flow problem  order 1 after scalings}. This would not be the case if, instead of circles, we had considered obstacles with fewer than two orthogonal planes of symmetry. Finally, to obtain a closed homogenized flow problem, we consider the $\order{\epsa}$ terms in (\ref{eq: Flow problem spatial transformed}b--d), given by
\begin{subequations}
\label{eq: flow order epsa}
\begin{alignat}{2}
\label{eq: cont order epsa}
\nabla_{\bs{y}} \bcdot \uu_1 + \nabla_{\bs{x}} \bcdot &\uu_0 = 0,& \qquad &\bs{y} \in \regf(\bs{x},t), \\
\label{eq: slip order epsa}
&\uu_1 =  - \alpha \pbyp{\rad}{t}\bs{n}_{\bs{y}} ,& \qquad &\bs{y} \in \partial \regf(\bs{x},t), \\
\label{eq: periodic flow order epsa}
&\uu_1 \text{ periodic},& \qquad &\bs{y} \in \partial \rege(\bs{x},t).
\end{alignat}
\end{subequations}
Integrating \eqref{eq: cont order epsa} over $\regf(\bs{x},t)$ and using Reynolds transport theorem in conjunction with the periodic and no-slip boundary conditions, (\ref{eq: flow order epsa}b,c) yields a continuity equation for the macroscale fluid velocity
\begin{align}
\label{eq: Incomp}
\nabla_{\bs{x}} \bcdot \UU = \alpha \dfrac{|\partial \regf(\bs{x},t)|}{|\rege (\bs{x},t)|} \pbyp{\rad}{t}.
\end{align}
The system \eqref{eq: Darcy velocity} and \eqref{eq: Incomp} determines the flow problem given the obstacle structure, which manifests itself through the source term on the right-hand side of \eqref{eq: Incomp}. This reflects the fact that reducing the available pore space within a cell for an incompressible fluid will push the fluid out of that cell. If $\partial \rad / \partial t \equiv 0$, we recover the system derived in \citet{dalwadi2015understanding}.

\subsection{Solute-transport problem}

Under the spatial transforms~\eqref{eq: spatial transform} and~\eqref{eq: normal in microscale}, the solute-transport problem~\eqref{eq: transport problem ND} becomes
\begin{subequations}
\label{eq: transport problem spatial transform}
\begin{alignat}{2}
\label{eq: transport problem spatial transform eq}
\epsa^2 \alpha \pbyp{\conc}{t} &= \left(\nabla_{\bs{y}} + \epsa \nabla_{\bs{x}}\right) \bcdot \left(\Pec^{-1} \left(\nabla_{\bs{y}} + \epsa \nabla_{\bs{x}} \right) \conc - \epsa \uu \conc \right),& \qquad &\bs{y} \in \regf(\bs{x},t), \\
\label{eq: transport problem spatial transform BC}
-\epsa^2 \kk \conc &= \left(\bs{n}_{\bs{y}} + \epsa \nabla_{\bs{x}} \rad \right) \bcdot\left( \Pec^{-1} \left(\nabla_{\bs{y}} + \epsa \nabla_{\bs{x}}\right) \conc  \right) + \order{\epsa^3},& \qquad &\bs{y} \in \partial \regf(\bs{x},t), \\
\conc &\text{ periodic},& \qquad &\bs{y} \in \partial \rege(\bs{x},t).
\end{alignat}
\end{subequations}
Expanding $\conc(\bs{x},\bs{y},t)$ in powers of $\epsa$, we find that that the leading-order concentration, $\conc_0$, is independent of the microscale, \emph{i.e.}, $\conc_0(\bs{x},t)$. The $\order{\epsa}$ terms in~\eqref{eq: transport problem spatial transform} allow us to write $\conc_1$ as
\begin{align}
\label{eq: conc_1 in terms of gamma}
\conc_1(\bs{x},\bs{y},t) = - \gam(\bs{x},\bs{y},t) \bcdot \nabla_{\bs{x}} \conc_0(\bs{x},t),
\end{align}
where the components of the function $\gam$ satisfy the cell problem
\begin{subequations}
\label{eq: cell transport problem}
\begin{alignat}{5}
0 &= \nabla_{\bs{y}}^2 \Gamma_i, & \qquad &\bs{y} \in \regf(\bs{x},t),\\
n_{y,i} &= \bs{n}_{\bs{y}} \bcdot \nabla_{\bs{y}} \Gamma_i, & \qquad &\bs{y} \in \partial \regf(\bs{x},t), \\
\Gamma_i & \text{ periodic},& \qquad &\bs{y} \in \partial \rege(\bs{x},t),
\end{alignat}
\end{subequations}
and $n_{y,i}$ is the $i{\text{th}}$ component of the unit vector $\bs{n}_{\bs{y}}$.

The effect of the obstacle growth only appears at the next order. Namely, the $\order{\epsa^2}$ terms in~\eqref{eq: transport problem spatial transform} are
\begin{subequations}
\label{eq: transport problem order delta^2}
\begin{alignat}{5}
\label{eq: transport problem order delta^2 P}
\alpha \pbyp{\conc_0}{t} &= \nabla_{\bs{y}} \bcdot \left( \Pec^{-1} \left( \nabla_{\bs{y}} \conc_2  + \nabla_{\bs{x}} \conc_1 \right) - \uu_1 \conc_0 - \uu_0 \conc_1 \right) \notag \\
&\quad + \nabla_{\bs{x}} \bcdot \left(\Pec^{-1}\left( \nabla_{\bs{y}} \conc_1  + \nabla_{\bs{x}} \conc_0 \right) - \uu_0 \conc_0 \right), & \qquad &\bs{y} \in \regf(\bs{x},t), \\
\label{eq: transport problem order delta^2 BC}
 - \kk \conc_0 &= \bs{n}_{\bs{y}} \bcdot \left( \Pec^{-1} \left( \nabla_{\bs{y}} \conc_2  + \nabla_{\bs{x}} \conc_1 \right) \right) \notag \\
&\quad + \nabla_{\bs{x}} \rad \bcdot \left( \Pec^{-1} \left(\nabla_{\bs{y}} \conc_1  + \nabla_{\bs{x}} \conc_0 \right)\right), & \qquad &\bs{y} \in \partial \regf(\bs{x},t), \\
\label{eq: transport problem order delta^2 periodic}
\conc_2 & \text{ periodic},& \qquad &\bs{y} \in \partial \rege(\bs{x},t).
\end{alignat}
\end{subequations}
Integrating~\eqref{eq: transport problem order delta^2 P} over $\regf$ and applying the boundary conditions \eqref{eq: Flow problem spatial transformed no slip}, \eqref{eq: slip order epsa}, and (\ref{eq: transport problem order delta^2}b--c) gives
\begin{align}
\int_{\regf(\bs{x},t)} \! \alpha \pbyp{\conc_0}{t} \, \mathrm{d}\bs{y} = & \int_{\regf(\bs{x},t)} \! \nabla_{\bs{x}} \bcdot \left(\Pec^{-1} \left(\nabla_{\bs{y}} \conc_1  + \nabla_{\bs{x}} \conc_0 \right)  - \uu_0 \conc_0 \right)\,\mathrm{d}\bs{y} \notag \\
& -\int_{\partial \regf(\bs{x},t)} \! \nabla_{\bs{x}} \rad \bcdot \left(\Pec^{-1} \left(\nabla_{\bs{y}} \conc_1  + \nabla_{\bs{x}} \conc_0 \right) - \uu_0 \conc_0 \right) \, \mathrm{d}s \notag \\
\label{eq: sec cond in terms of integrals pre-RTT}
& - \int_{\partial \regf(\bs{x},t)} \! \kk \conc_0 \, \mathrm{d}s + \int_{\partial \regf(\bs{x},t)} \! \alpha \pbyp{\rad}{t}  \conc_0 \, \mathrm{d}s,
\end{align}
where $\mathrm{d}s$ denotes the differential element of the obstacle surface $\partial \regf(\bs{x},t)$. Using Reynolds transport theorem, \eqref{eq: sec cond in terms of integrals pre-RTT} reduces to 
\begin{align}
\label{eq: sec cond in terms of integrals}
\alpha \pbyp{}{t} \left(|{\regf}({\bs x},t)| \conc_0 \right) = \nabla_{\bs{x}} \bcdot \int_{\regf(\bs{x}, t)} \! \left( \Pec^{-1}  \left(\nabla_{\bs{y}} \conc_1  + \nabla_{\bs{x}} \conc_0 \right) - \uu_0 \conc_0 \right) \, \mathrm{d}\bs{y} - \int_{\partial \regf(\bs{x},t)} \! \kk \conc_0 \, \mathrm{d}s,
\end{align}
noting that $c_0$ is independent of $\bs y$. From now on, we simply write $\porosity$ but it should be understood that the porosity will be, in general, a function of $\bs x$ and $t$. In a similar manner, henceforth we use $\regf(\porosity)$ instead of $\regf(\bs{x},t)$ and likewise for $\partial \regf$. Finally, using \eqref{eq: conc_1 in terms of gamma}, \eqref{eq: sec cond in terms of integrals} reduces to
\begin{align}
\label{eq: effective intrinsic equation}
\alpha \pbyp{}{t} \left(|\regf(\porosity)| \conc_0 \right) &= \nabla_{\bs{x}} \bcdot \left(\Pec^{-1}  \left(\int_{\regf(\porosity)} ( \Iu - \! \Jt) \, \mathrm{d}\bs{y}\right) \nabla_{\bs{x}} \conc_0 
\hspace{-0.1mm}- \hspace{-0.5mm}\left(\int_{\regf(\porosity)}  \uu_0 \, \mathrm{d}\bs{y} \right) \conc_0 \right) \hspace{-0.2mm} \notag \\
& \quad- \hspace{-0.2mm}|\partial \regf(\porosity)| \kk \conc_0,
\end{align}
where $(\Jt)_{ij} = \partial \Gamma_j / \partial y_i$ is the transpose of the Jacobian matrix of $\gam$, the solution to the cell problem \eqref{eq: cell transport problem}. 

To express \eqref{eq: effective intrinsic equation} in terms of the volumetric average concentration $\creal ({\bs x},t)$ defined in \eqref{eq: Volumetric average concentration}, we note that $\creal({\bs x},t) \sim |\rege|^{-1} \int_{\regf(\porosity)} \conc_0 \mathrm{d}\bs{y} = \porosity \conc_0({\bs x},t)$ at leading order in $\epsa$. Using this relation and \eqref{eq: Darcy velocity order 1}, we find that
\begin{subequations}
\label{eq: effective final problem}
\begin{align}
\label{eq: effective volumetric equation}
\alpha \pbyp{\creal}{t} = \nabla_{\bs{x}} \bcdot \left(\Deff(\porosity) \nabla_{\bs{x}} \creal - \dfrac{\creal}{\porosity}\left(\UU(\porosity) + \Deff(\porosity) \nabla_{\bs{x}} \porosity \right)\right) - f(\porosity) \creal,
\end{align}
at leading order in $\epsa$, where the effective diffusion coefficient is
\begin{align}
\label{eq: effective coefficient matrices}
\Deff(\porosity) \Iu = \Pec^{-1} \left( \Iu - \dfrac{1}{|\regf(\porosity)|}\int_{\regf(\porosity)} \! \Jt \, \mathrm{d}\bs{y} \right),
\end{align}
and the effective adsorption coefficient is 
\begin{align}
\label{eq: particular sink function}
f(\porosity) = \kk \dfrac{|\partial \regf(\porosity)|}{|\regf(\porosity)|} = \dfrac{4 \kk \pi \rad}{\porosity \sqrt{3}},
\end{align}
using the fact that $\porosity = 1 - 2 \pi \rad^2 / \sqrt{3}$, and $|\partial \regf(\porosity)| = 4 \pi \rad$. As is the case for the permeability $\mathcal{K}$, we find that the diffusion tensor is a multiple of the identity due to the symmetry of the cell problem~\eqref{eq: cell transport problem}.
Finally, the condition to couple contaminant concentration with obstacle growth \eqref{eq: transport problem ND growing obstacle} becomes
\begin{align}
\label{eq: Couple cont with growth full}
\porosity \pbyp{\rad}{t} = \creal.
\end{align}
\end{subequations}

\begin{figure}
\centering
    	\includegraphics[width=\textwidth]{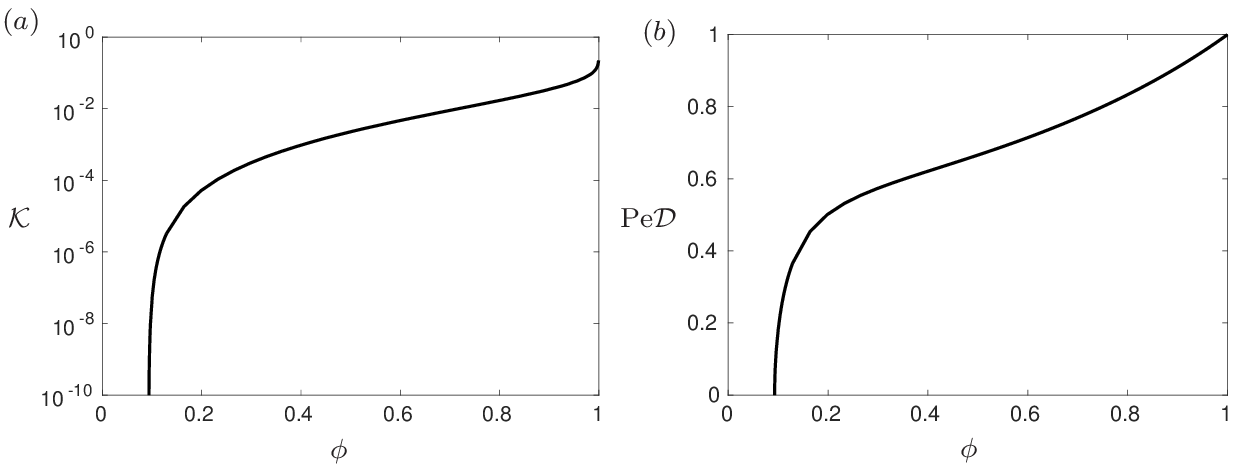}
\caption{The functions (a) $\mathcal{K}(\porosity)$ and (b) $\Pec \Deff(\porosity)$ (defined in \eqref{eq: def of K} and \eqref{eq: effective coefficient matrices} respectively) calculated using Comsol Multiphysics, for circular obstacles whose centres lie on a hexagonal lattice.}
\label{fig: Perm and Diff coeff}
\end{figure}

The effective permeability $\mathcal{K}(\porosity)$ (from \eqref{eq: def of K}) and the effective diffusion $\Deff(\porosity)$ (from \eqref{eq: effective coefficient matrices}) that encapsulate the microscale physics can be computed by solving the respective cell problems \eqref{eq: Flow problem  order 1 after scalings} and \eqref{eq: cell transport problem} for a given cell porosity $\phi$, determined by the size of the obstacles. We calculate this numerically using the finite-element software Comsol Multiphysics and find that $\mathcal{K}$ and $\Deff$ are both monotonically increasing with the porosity, as expected, and that there is a sharp decrease in both functions as the porosity tends down towards the critical porosity where blocking occurs (see figure~\ref{fig: Perm and Diff coeff}).

The homogenization procedure has significantly reduced the mathematical complexity of the growing multiply-connected domain in the full problem~\eqref{eq: transport problem ND}, with only a marginal increase in the complexity of the coefficients in the resulting governing equations~\eqref{eq: effective final problem}, which capture the microscale structure of the problem in a systematic manner.

\section{A unidirectionally graded filter}
\label{sec: Uni-dir problem}
\subsection{Model set-up}
\label{sec: Model set-up}

We are now in a position to use the homogenized equations \eqref{eq: Darcy velocity}, \eqref{eq: Incomp}, and \eqref{eq: effective final problem} to quantify the effect of blocking and porosity gradients on filter efficiency. We consider a canonical industrial process of interest whereby a three-dimensional filter separates two reservoirs, and a unidirectional flow is induced through the filter. As is common in industrial set-ups, we assume that the filter is graded in the same direction as the fluid flow. This reduces the problem to a one-dimensional description. 

We define the direction of porosity variation as $x$, where $x \in (0, 1)$ within the filter (since the dimensional characteristic length $\leng$ is chosen to be the depth of the filter). Thus the set-up is similar to that illustrated in figure~\ref{fig: Set-up schematic}. The upstream is defined for $x \in (-\infty,0)$ and the downstream for $x \in (1,\infty)$.

Provided the boundary conditions allow for unidirectionality, equations \eqref{eq: Darcy velocity} and \eqref{eq: Incomp} yield a unidirectional macroscale flux $\UU = U(x,t) \bs{e}_x$ (where $\bs{e}_x$ is the unit vector in the $x$-direction). When filtering at constant (dimensionless) pressure drop $\pdiff$, we find that 
\begin{multline}
\label{eq: 1D flow}
U(x,t) = \Kav(t) \left(\pdiff - \alpha \int_0^1 \! \frac{1}{\mathcal{K}(\phi(v,t))}\lr{\int_0^v \! |\partial \regf(u,t)| \pbyp{\rad}{t}(u,t) \, \mathrm{d}u}  \, \mathrm{d}v \right) \\+ \alpha \int_0^x \! |\partial \regf(\xi,t)| \pbyp{\rad}{t}(\xi,t) \, \mathrm{d} \xi,
\end{multline}
where
\begin{align} \label{Kbar}
\Kav(t) = \left(\int_0^1 \dfrac{\mathrm{d}x}{\mathcal{K}(x,t)} \right)^{-1}.
\end{align}

Thus the governing equation \eqref{eq: effective volumetric equation} becomes
\begin{align}
\label{eq: example governing equation}
\alpha \pbyp{\creal}{t} = \pbyp{}{x} \left[\Deff(\porosity) \pbyp{\creal}{x} - \dfrac{\creal}{\porosity}\left(U(x,t) + \Deff(\porosity) \pbyp{\porosity}{x} \right)\right] - f(\porosity) \creal, \qquad x \in (0,1),
\end{align}
and the relationship between obstacle growth and concentration, ~\eqref{eq: Couple cont with growth full}, is
\begin{align}
\label{eq: Constitutive law 1D}
\porosity \pbyp{\rad}{t} = \creal, \qquad x \in (0,1) .
\end{align}

Considering the limit of \eqref{eq: effective coefficient matrices} and \eqref{eq: example governing equation} as $\porosity \to 1$, and using continuity of fluid flux, we obtain the following governing equations for the upstream $\cus$ and downstream $\cds$ concentrations
\begin{subequations}
\label{eq: additional systems}
\begin{alignat}{2}
\label{eq: additional systems a}
\alpha \pbyp{\cus}{t} &= \pbyp{}{x} \left(\Pec^{-1} \pbyp{\cus}{x} - U(0,t) \cus \right),& \qquad &x \in (-\infty,0), \\
\label{eq: additional systems b}
\alpha \pbyp{\cds}{t} &= \pbyp{}{x} \left(\Pec^{-1} \pbyp{\cds}{x} - U(1,t) \cds\right),& \qquad &x \in (1,\infty).
\end{alignat}
\end{subequations}

We impose continuity of concentration and concentration flux at the boundaries between the filter and the reservoirs. In the far-field of the reservoirs, the concentration tends to a constant value. We may take the upstream concentration $\cus \to 1$ (by choice of our non-dimensionalization), while the downstream concentration $\cds$ tends to a constant value that must be determined as part of the solution. Mathematically, this corresponds to
\begin{subequations}
\label{eq: additional systems BC}
\begin{alignat}{2}
\label{eq: x to minus inf BC}
\cus &\to 1, &\qquad &x \to -\infty, \\
\label{eq: cont of c x=0}
\cus &= \dfrac{\creal}{\porosity}, &\qquad &x =  0, \\
\Pec^{-1} \pbyp{\cus}{x} - U \cus &= \Deff(\porosity) \pbyp{\creal}{x} - \dfrac{\creal}{\porosity}\left(U + \Deff(\porosity) \pbyp{\porosity}{x} \right), &\qquad &x =  0, \\
\label{eq: DS conc}
\cds &= \dfrac{\creal}{\porosity}, &\qquad &x =  1, \\
\label{eq: cont of flux x=L}
\Pec^{-1} \pbyp{\cds}{x} - U \cds &= \Deff(\porosity) \pbyp{\creal}{x} - \dfrac{\creal}{\porosity}\left(U + \Deff(\porosity) \pbyp{\porosity}{x} \right), &\qquad &x =  1, \\
\label{eq: x to inf BC}
\dbyd{\cds}{x} &\to 0, &\qquad & x \to \infty.
\end{alignat}
\end{subequations}
Thus, with appropriate initial conditions, our homogenized system is given by \eqref{eq: 1D flow}--\eqref{eq: additional systems BC}. We solve this system numerically using the method of lines, discretizing in space with a second-order finite difference scheme and using the MATLAB program \texttt{ode15s} to solve in time. 

A suitable measure of filter efficiency is the cumulative contaminant removal, defined by
\begin{align}
\label{eq: cumulative adsorption}
\cumadsorp(t) := \int_0^1 \! \left(\porosity(x,0) - \porosity(x,t)\right) \, \mathrm{d}x = \dfrac{1}{\kk}\int_0^t \! \int_0^1 \! f(\porosity(x,s)) \creal(x,s) \, \mathrm{d}x   \, \mathrm{d}s,
\end{align}
where the second equality can be deduced from using the relationship $\porosity = 1 - 2 \pi \rad^2 / \sqrt{3}$ to obtain $\partial \porosity / \partial t = -(|\partial \regf|/ |\rege|) \partial \rad / \partial t $, in conjunction with \eqref{eq: particular sink function} and \eqref{eq: Constitutive law 1D}. A measure of the corresponding structural changes within the filter is captured by the quantity 
\begin{align}
\label{eq:pore size}
P(x,t)=1-2\rad(x,t)
\end{align}
which expresses the ratio of the distance between obstacles and the obstacle centres, or an effective pore size in the filter.

We begin by considering the case of a filter whose initial porosity is uniform. Solving \eqref{eq: additional systems BC} we find that as filtration proceeds the pore size $P(x,t)$ decreases in time throughout the filter (figure \ref{fig: compare_num_v_asy}a) and the rate of contaminant removal $\mathcal{T}'(t)$ reduces (figure~\ref{fig: compare_num_v_asy}b). The reduction in pore size is more rapid close to the filter entrance than towards the filter exit (figure~\ref{fig: compare_num_v_asy}a), which causes the filter to block at the entrance while the rest of the filter is still fit for purpose. This has significant practical disadvantages, and so we now use our model to explore strategies that minimize this wastefulness by varying the initial filter porosity. To do so, we first note that the homogenized system \eqref{eq: additional systems BC} can be further simplified by exploiting two relevant asymptotic regimes. We discuss these in the following section.

\subsection{Asymptotic reductions of the homogenized equations}
\label{sec: Asymptotic reductions}

In the majority of filtration applications the growth of the obstacles, and thus the blocking of the filter, takes place over a timescale that is much longer than the time taken for the fluid to flow through the filter. Mathematically, this corresponds to the quasi-steady limit in which $\alpha \to 0$. At leading order in this limit, returning to the three-dimensional equations for generality, the flow equations \eqref{eq: Darcy velocity}, \eqref{eq: Incomp} become
\begin{subequations}
\label{eq: Asymptotic equations}
\begin{align}
\UU({\bs x},t) = -\mathcal{K}(\porosity) \nabla_{\bs{x}} p, \quad \nabla_{\bs{x}} \bcdot \UU = 0,
\end{align}
where $\mathcal{K}(\porosity)$ is defined in \eqref{eq: def of K}; and the transport equations~\eqref{eq: effective final problem} become
\begin{align}
\nabla_{\bs{x}} \bcdot \left(\Deff(\porosity) \nabla_{\bs{x}} \creal - \dfrac{\creal}{\porosity}\left(\UU(\porosity) + \Deff(\porosity) \nabla_{\bs{x}} \porosity \right)\right) = f(\porosity) \creal, \quad \porosity \pbyp{\rad}{t} = \creal,
\end{align}
\end{subequations}
where $\Deff(\porosity)$ and $f(\porosity)$ are defined in (\ref{eq: effective final problem}b,c), respectively. In this regime, the growth in $\rad$ decouples from the flow and transport problems, which are now quasi-steady.

For a unidirectional porosity variation, the flow velocity becomes independent of $x$ and so (\ref{eq: Asymptotic equations}a) simplifies to
\begin{align}
\label{eq:U asymptotic 1}
U(t) = \Kav(t) \pdiff,
\end{align}
where $\Kav$ is given in \eqref{Kbar}. Moreover, the upstream and downstream concentrations can be solved to obtain
\begin{align}
\label{eq: up/downstream conc}
\cus = 1 - A(t) \exp(\Pec \, U(t) x),  \quad \cds = B(t),
\end{align}
where $A(t)$ and $B(t)$ are arbitrary functions of time. Substituting \eqref{eq: up/downstream conc} into \eqref{eq: additional systems BC}, the system for $\creal$ becomes
\begin{subequations}
\label{eq: reduced conc gov eq}
\begin{alignat}{2}
\label{eq: reduced conc gov eq EQ}
f(\porosity) \creal &= \pbyp{}{x} \left[\Deff(\porosity) \pbyp{\creal}{x} - \dfrac{\creal}{\porosity}\left(U(t) + \Deff(\porosity) \pbyp{\porosity}{x} \right)\right], &\qquad &x \in (0,1), \\
- U(t) &= \Deff(\porosity) \pbyp{\creal}{x} - \dfrac{\creal}{\porosity}\left(U(t) + \Deff(\porosity) \pbyp{\porosity}{x} \right), &\qquad &x =  0, \\
0 &= \pbyp{\creal}{x} - \dfrac{\creal}{\porosity} \pbyp{\porosity}{x}, &\qquad &x =  1.
\end{alignat}
The governing equation for the porosity \eqref{eq: Constitutive law 1D} is unchanged:
\begin{align}
\label{eq: Constitutive law 1D again}
\porosity \pbyp{\rad}{t} = \creal, \qquad x \in (0,1) .
\end{align}
\end{subequations}
This problem is similar to that considered in \citet{dalwadi2015understanding}, where the absence of microstructural changes due to filter clogging meant that $U(t) \equiv 1$ in \eqref{eq: reduced conc gov eq}, and $\partial \rad / \partial t \equiv 0$ instead of \eqref{eq: Constitutive law 1D again}. We can solve the system \eqref{eq: reduced conc gov eq} numerically using the method of lines, discretizing in space with a second-order finite difference scheme and using the MATLAB program \texttt{ode15s} to solve in time. 

We can make further analytic progress by considering the case where, in addition to the slow obstacle growth, the deviation of the porosity $\porosity(x,t)$ from its average value in space $\pav(t) = \int_0^1 \! \porosity(x,t) \, \mathrm{d}x$ is small. In this case, we express the porosity as its average in space plus the deviation, as follows $\porosity(x,t) = \pav(t) + \pc(x,t)$, and assume that $|\pc| \ll \pav$. Thus, the leading-order coefficients in the linear governing equation \eqref{eq: reduced conc gov eq EQ} are constant in $x$, and the solution can be written in terms of hyperbolic functions. The $\order{\pc}$ correction terms in \eqref{eq: reduced conc gov eq EQ} can then be solved using the method of variation of parameters \citep{dalwadi2015understanding}. From this, we may express the concentration $\creal$ as a sum of the concentration distribution in a filter with constant porosity, $\creal_0(x,t)$, and the adjustment due to the porosity gradient, $\creal_1(x,t)$, \emph{i.e.},
\begin{align}
\label{eq: asymptotic concentration}
\creal(x,t) \sim \creal_0(x,t) + \creal_1(x,t),
\end{align}
where $|\creal_1| \ll \creal_0$; $\creal_0$ and $\creal_1$ are functions of $\porosity$ that can be solved for explicitly and are given in Appendix~\ref{App: asymptotic coefficients}. Using this analytic solution, we are able to reduce the entire problem to solving \eqref{eq: Constitutive law 1D again}. We solve this system numerically using the method of lines, discretizing in space with a second-order finite difference scheme and using the MATLAB program \texttt{ode15s} to solve in time. 

\begin{figure}
\centering
    	\includegraphics[width=\textwidth]{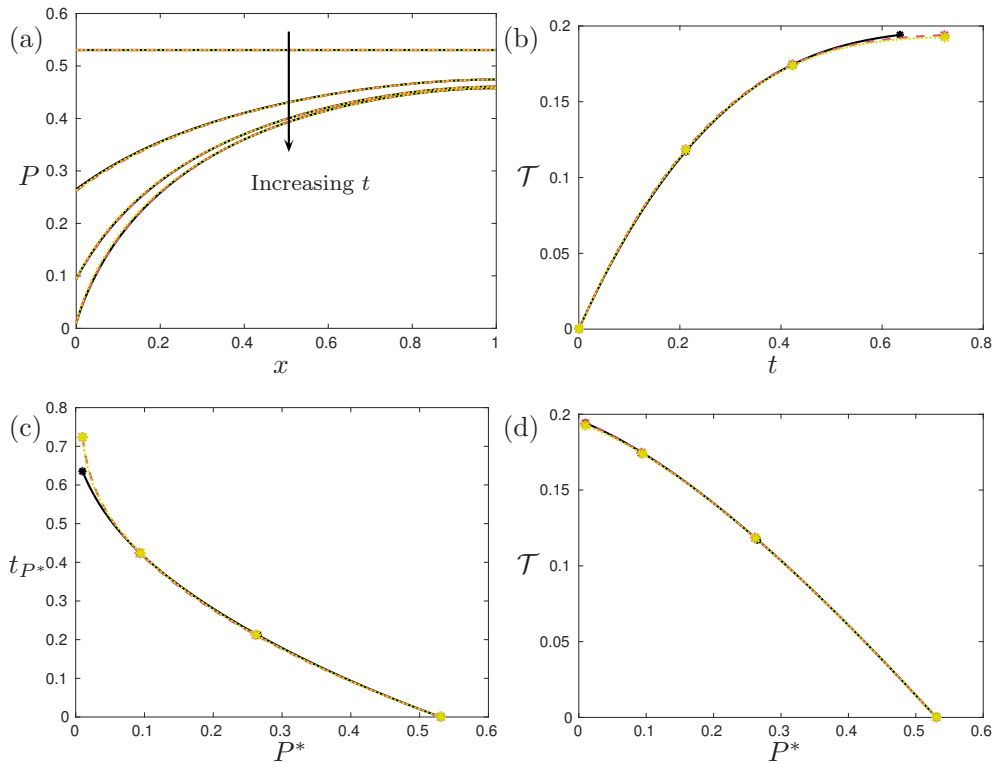}
\caption{(Colour online) Filtration through an initially uniform porosity filter, $\porosity(x,0) = 0.8$, with $\Pec = 3$, $\kk = 1$. (a) The pore size, $P$, defined by \eqref{eq:pore size}, throughout the filter at four snapshots in time (shown by asterisks in (b), (c), and (d)). We start with $t = 0$ and end with the first time at which a pore size in the filter reaches $0.01$, defined as $t_{0.01}$. The two intermediate snapshots are at time $t_{0.01}/3$ and $2 t_{0.01}/3$, where we use $t_{0.01}$ calculated from the solution to the full homogenized equations for all cases. (b) The cumulative contaminant adsorption $\cumadsorp$, defined by (\ref{eq: cumulative adsorption}), as a function of time. (c) The time taken until the pore size reaches a minimum value. (d) The cumulative contaminant adsorption $\cumadsorp$, defined by (\ref{eq: cumulative adsorption}), as a function of minimum pore size. The solid black curves denote the solutions to the full homogenized equations \eqref{eq: 1D flow}--\eqref{eq: additional systems BC}, the dashed red curves denote the asymptotic approximation when obstacle growth is small, \eqref{eq: reduced conc gov eq}, and the dotted yellow curves denote the asymptotic solution when we further assume a weak porosity gradient, \eqref{eq: asymptotic concentration}. We see excellent agreement throughout, with only small deviations that occur when the pore size becomes small.}
\label{fig: compare_num_v_asy}
\end{figure}

In figure \ref{fig: compare_num_v_asy}, we compare the results from these two asymptotic results, \eqref{eq: reduced conc gov eq} and \eqref{eq: asymptotic concentration}, with the full numerical solution, \eqref{eq: 1D flow}--\eqref{eq: additional systems BC} (solid black lines), as dashed red and dotted yellow lines, respectively. Both asymptotic solutions offer very good agreement in the pore size $P$ with the full numerical solution (figure \ref{fig: compare_num_v_asy}a), and in $\cumadsorp$ until around $t=0.45$ (figure \ref{fig: compare_num_v_asy}b). 

The discrepancy between numeric and asymptotic solutions appears when considering the time taken to reach a minimum pore size rather than in the cumulative contaminant removal metric $\cumadsorp$. We are able to contextualize this metric by introducing relative timings in a filter lifespan. That is, we define the minimum pore size
\begin{align}
P^*(t) = \min_{x \in (0, 1), \, t \geq 0} P(x,t),
\end{align}
and the time taken until the minimum pore size reaches a given value,
\begin{align}
t_{P^*} = \min \left\{ t \geq 0 : P(x,t) \leq P^*, \, \text{for some } x \in (0,1)\right\}.
\end{align}
Blocking occurs at $t_{0}$, but we must consider small finite values of $P^*$ due to the singularity in \eqref{eq: reduced conc gov eq} when $\porosity \to 0$, and hence when $P^* \to 0$, at the point of blocking. For example, in figure \ref{fig: compare_num_v_asy} the simulations are stopped at $t_{0.01}$.

From figure \ref{fig: compare_num_v_asy}b, we see that although the maximum time to removal is different between the asymptotic and numerical solutions, the total cumulative removals are very close. We show the former in more detail in figure \ref{fig: compare_num_v_asy}c, where we see that the numerical and asymptotic solutions agree well in $t_{P^*}$ until $P^*$ falls below around $P^* = 0.05$. We show the latter in more detail in figure \ref{fig: compare_num_v_asy}d, where we use $P^*$ as the abscissa instead of time, against cumulative contaminant removal. We see that there is excellent agreement between numeric and asymptotic solutions throughout, and henceforth use the full asymptotic approximation \eqref{eq: asymptotic concentration}, which offers a very accurate representation of the solution to the full homogenized system \eqref{eq: 1D flow}--\eqref{eq: additional systems BC} while enabling further analytic study of the system. Moreover, we note that we halt simulations when $P^*$ reaches $0.01$, and refer to this as `blocking', and we see from figure \ref{fig: compare_num_v_asy}d that this is unlikely to yield major differences in values of total contaminant removal.

\subsection{Linearly graded filters}
\label{sec: Model analysis}

As shown in figure \ref{fig: compare_num_v_asy}, in a depth filter with a uniform porosity the early portion of the media will block before the later portion has been used to its full effect. This can be counteracted by using a filter whose porosity decreases with depth.

\begin{figure}
\centering
	\includegraphics[width=\textwidth]{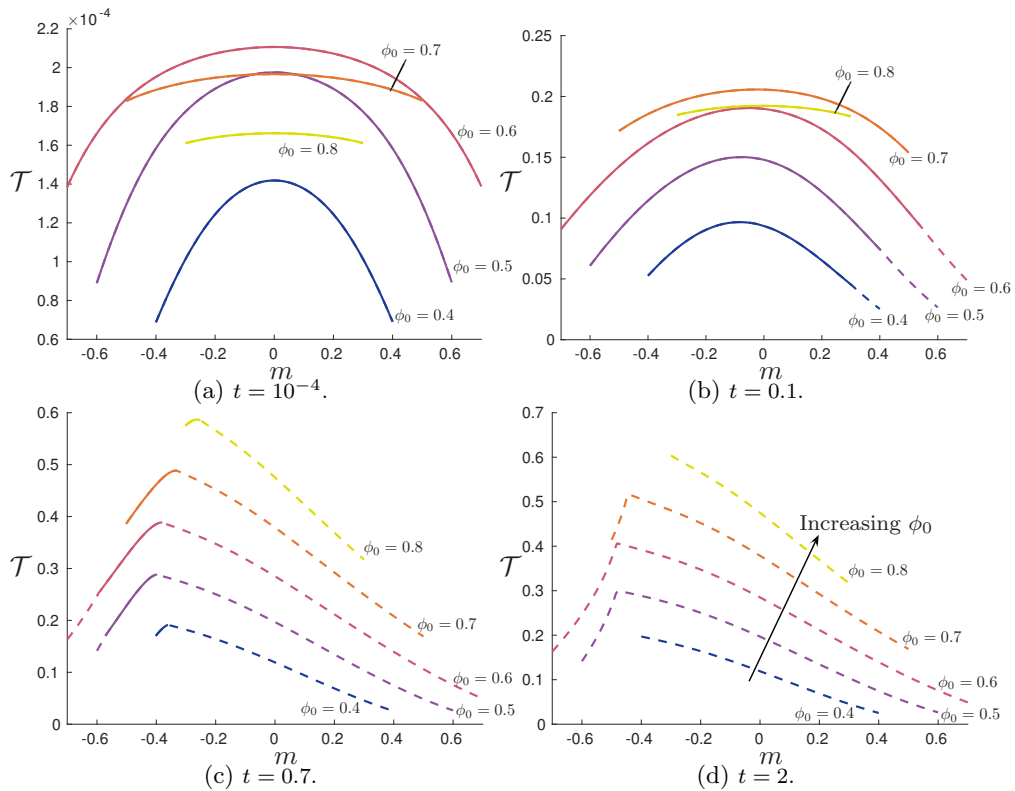}
\caption{(Colour online) Cumulative contaminant removal, $\cumadsorp$, defined by (\ref{eq: cumulative adsorption}), for varying $\porosity_0$ and $m$ in initial porosity distributions of the form $\porosity(x) = \porosity_0 + m(x - 0.5)$, given by the solution to \eqref{eq: 1D flow}--\eqref{eq: additional systems BC}. Each curve represents a different value of $\porosity_0$, which increments in steps of $0.1$ from $\porosity_0 = 0.4$ to $\porosity_0 = 0.8$ and the arrow in (d) denotes curves of increasing $\porosity_0$. We vary $m$ for a given $\porosity_0$ such that $\porosity \in [0.2,0.95]$. Therefore, the available range of $m$ varies with $\porosity_0$. We use the reference values $\porosity = 0.6$, $\Pec = 5$, $\kk = 0.1$ from which to modify appropriate parameters. The dashed curves denote that the minimum pore size has reached $0.01$ and the filtration has stopped.}
\label{fig: T comparisons}
\end{figure}

\begin{figure}
\centering
\includegraphics[width=\textwidth]{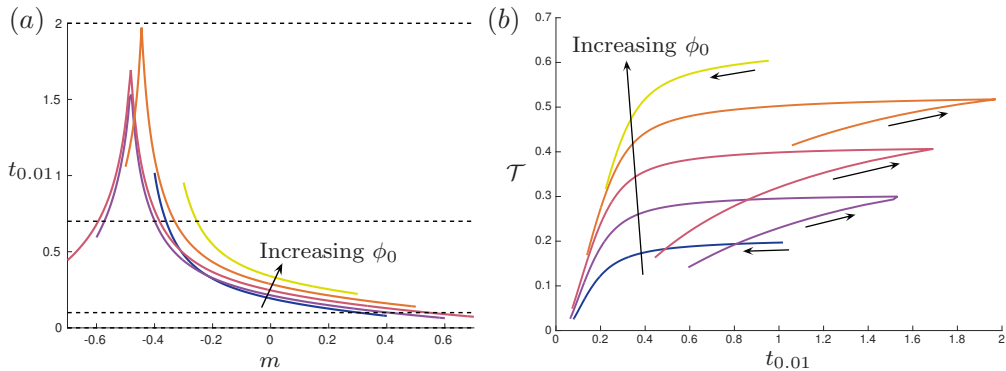}
\caption{(Colour online) (a) The time until the minimum pore size reaches 0.01 for varying $\porosity_0$ and $m$ in initial porosity distributions of the form $\porosity(x) = \porosity_0 + m(x - 0.5)$, determined by solving \eqref{eq: 1D flow}--\eqref{eq: additional systems BC}. The dashed black lines denote the times at which snapshots are taken in figures \ref{fig: T comparisons} and \ref{fig: S comparisons}. (b) The total cumulative adsorption over the lifespan of a filter versus the time taken to block. Each line represents a different initial average porosity, and the unlabelled arrows denote the direction of increasing $m$ for each line. In both figures, each line represents a different value of $\porosity_0$, which increments in steps of $0.1$ from $\porosity_0 = 0.4$ to $\porosity_0 = 0.8$ and the labelled arrow denotes lines of increasing $\porosity_0$. We vary $m$ for a given $\porosity_0$ such that $\porosity \in [0.2,0.95]$. Therefore, the available range of $m$ varies with $\porosity_0$. We use the reference values $\porosity = 0.6$, $\Pec = 5$, $\kk = 0.1$ from which to modify appropriate parameters. }
\label{fig: Time until blocking}
\end{figure}

In this section we consider linearly graded filters at $t=0$, characterized by $\porosity(0,x) = \porosity_0 + m(x - 0.5)$, where we vary the initial average porosity, $\porosity_0$, and the initial porosity gradient, $m$. In a similar manner to \citet{dalwadi2015understanding},  we assume that the pressure drop is kept constant across each filter considered and, noting that the system is invariant to the scaling
\begin{align}
\label{eq: invariance}
(U(0), \Pec, \kk) \mapsto (\omega U(0), \Pec/\omega, \omega \kk),
\end{align}
for any constant $\omega$, we impose $U(0) = 1$ and vary $\Pec$ and $\kk$ accordingly. For short times, a filter with spatially uniform porosity (\emph{i.e.} $m=0$) provides the largest cumulative removal for a given average porosity, and there is an optimum value for this average porosity (figure~\ref{fig: T comparisons}a). This optimum exists because a highly porous filter has less surface area with which to remove contaminant, but a less porous filter will admit a lower flow rate through the filter, relying more on diffusion than advection for contaminant transport, and thus lowering the contaminant removal. The result that there is no apparent difference in the total contaminant removal of a filter with a positive or negative porosity gradient for small time replicates the results for the initial instantaneous adsorption obtained in \citet{dalwadi2015understanding}. However, in contrast to \cite{dalwadi2015understanding} we are now able to capture the asymmetry in removal that arises over longer times as a result of blockages within the filter. In particular, negative initial porosity gradients do provide an inherently larger cumulative removal at a given time than positive initial porosity gradients (figure~\ref{fig: T comparisons}b,c), and their final cumulative removal is larger (figure~\ref{fig: T comparisons}d). We note that whilst negative initial porosity gradients do last for a longer time before blocking, the filters that take the longest time to block do not automatically remove the most contaminant (figure~\ref{fig: Time until blocking}).

In \citet{dalwadi2015understanding}, a further metric was introduced to measure the uniformity of uptake, thus allowing us to account for the superior performance of filters with a negative porosity gradient without explicitly accounting for filter evolution. As we are able to account for filter evolution in this work, $\cumadsorp$ is now the better measure of filtration. We show in Appendix~\ref{sec: Uniformity of removal} that this alternative metric can only provide qualitative insights into filtration, and should not be used when the filter structure evolves in time.

For a given initial average porosity, the largest cumulative removal does not necessarily correspond to a filter with the most negative initial porosity gradient (figure~\ref{fig: T comparisons}d). If the porosity gradient is too negative, the removal is skewed towards the filter exit, and thus blocking occurs at the filter exit too quickly. This effect causes the sharp drop-off in filter performance seen in figures~\ref{fig: T comparisons}c,d for large negative values of $m$. Thus, an optimum initial porosity gradient exists for a given initial average porosity. Whilst this `optimal' initial porosity gradient maximizes contaminant removal over the space of initially \emph{linear graded} filters with a given initial average porosity, it will not necessarily maximize contaminant removal over the space of \emph{all} filters with a given initial average porosity. We explore this idea in more detail in the next section.

\subsection{Filters that block everywhere at once}

Using the framework we have set up in this paper, we can address the question of which initial porosity distributions maximize the cumulative removal for given operating conditions. We are able to bypass much of the difficulty involved in the full optimal control problem by noting that, for a given initial average filter porosity, the cumulative adsorption \eqref{eq: cumulative adsorption} is maximized when a filter blocks all at once, rather than in just one place. Most filters will only block in one place, and this is the case for all of the filters represented in figure~\ref{fig: T comparisons}, where the filters block at either the filter entrance or the filter exit.

By starting with a filter that is blocked everywhere and running our simulations backwards in time, we can obtain initial porosity distributions whose lifespans end with blocking occurring everywhere in the filter at once. To avoid the computational issues arising when blocking occurs, we approximate the concept of a filter that blocks everywhere at once by a filter that reaches a pore size of $0.01$ everywhere at once. An issue with this procedure is that the full homogenized system for the concentration \eqref{eq: 1D flow}--\eqref{eq: additional systems BC} is parabolic in time, and thus running the simulations backwards in time results in an ill-posed problem. However, the evolution equation for the filter porosity \eqref{eq: Constitutive law 1D again} is well-posed if run backwards in time, and only relies on knowing the concentration for a given porosity. Thus, if we use the asymptotic solutions for the concentration derived in \S\ref{sec: Asymptotic reductions}, we do not have to run the concentration evolution equation backwards in time and we are able to side-step the issue with the ill-posedness of this problem.

A related issue, due to solving the system of equations backwards in time, is that only the \emph{final} flow velocity can be imposed, and the \emph{initial} flow velocity is now an output to the calculation. Thus, a general final flow velocity $U(t_{0.01})$ will not necessarily result in $U(0) = 1$, the condition we prescribe in the forward problem. Moreover, we vary $\Pec$ and $\kk$ according to the \emph{initial} porosity in the forward problem, using the invariant scaling \eqref{eq: invariance}, and thus it is not immediately clear what values of $\Pec$ and $\kk$ we should impose for the inverse problem. We overcome these problems by exploiting the fact that we impose a constant pressure drop across the filter, and thus $\Kav(0) = U(0) \Kav(t_{0.01})/ U(t_{0.01}) = \Kav(t_{0.01})/ U(t_{0.01})$, where the first equality holds from \eqref{eq:U asymptotic 1}, and the second equality holds by using $U(0) = 1$ (the filter for which we are aiming). This allows us to move between the initial and final average permeabilities as a function of the final flow velocity, as well as imposing values of $\Pec$ and $\kk$ using the final porosity. Hence, we are able to use a shooting method for just one parameter, the final flow velocity $U(t_{0.01})$, shooting to achieve $U(0) = 1$.

To summarize, we impose a final constant pore size with $P(x,t_{0.01}) \equiv 0.01$ and a final flow velocity and run our simulation in reverse, halting the process when the porosity at any point reaches a set maximum. We iterate this process, varying the final flow velocity, until the initial flow velocity $U(0) = 1$ to within an accuracy of $2 \times 10^{-3}$. We repeat this process for different maximum porosity constraints to obtain a family of initial porosity distributions that block everywhere at once when contaminant flows through, each with a different maximum (and average) porosity. The maximum porosities we choose are given by $1 - \pi R^2$, where $R$ takes the values $0.1174$, $0.15$, $0.2$, $0.25$, and $0.3$. We produce this range of initial porosity distributions as there may be physical constraints on how large/small the porosity/fibres can be.

\begin{figure}
\centering
\includegraphics[width=\textwidth]{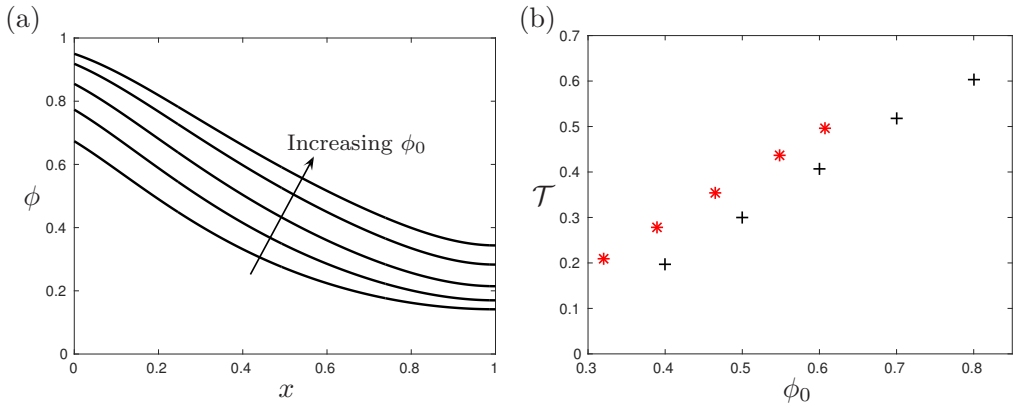}
\caption{(Colour online) (a) The initial porosity distributions for filters which block everywhere at once with different initial average porosities, $\phi_0$. The initial average porosities are: 0.320, 0.390, 0.465, 0.548, 0.607 (all to three significant figures). (b) The cumulative contaminant removal $\cumadsorp$ for the pseudo-optimal filters (red asterisks) compared to the linear filters considered in \S\ref{sec: Model analysis} (black crosses). The data for the black crosses is from figure~\ref{fig: T comparisons}. We use dimensionless parameters corresponding to the reference values $\porosity(x,0) \equiv 0.6$, $\Pec = 5$, $\kk = 0.1$ in the manner described in the text. In (b) we see that the pseudo-optimal filters do provide increased contaminant removal for a given initial average porosity, but the average porosity of the pseudo-optimal filters cannot be increased to an arbitrarily large value.}
\label{fig: ideal filter}
\end{figure}

Each resulting filter has a decreasing porosity with depth and, compared to the region near the filter entrance, the porosity gradient in each filter is more negative towards the centre of the filter and smaller towards the end of the filter (figure~\ref{fig: ideal filter}a). This effect is more pronounced for larger maximum porosities. We compare how these filters remove contaminant against the linear filters we considered in \S\ref{sec: Model analysis} and see that the ideal filters do remove around $20\%$ more contaminant for a given initial average porosity (figure~\ref{fig: ideal filter}b). However, these filters cannot have arbitrarily large initial average porosity, as a filter whose initial porosity is everywhere close to $1$ will not block everywhere at once, and thus a linearly graded filter with a large enough initial average porosity can remove more contaminant than a filter with a lower initial average porosity that blocks everywhere at once. As a result, we may term these \emph{pseudo-optimal} filters. As a filter may require a maximum porosity to maintain structural integrity, the choice of initial porosity distribution will depend on the operating conditions.

\section{Discussion}

\label{sec: Discussion}

We have systematically derived a macroscopic model for the dynamic blocking of a porosity-graded filter from microscale information by a generalization of standard homogenization theory for near-periodic systems. The result is an advection--diffusion--reaction equation for the solute concentration within the filter, a modified Darcy Law for the fluid transport, and an evolution equation for the underlying porosity of the filter.  This system depends on the initial porosity distribution of the filter and the operating conditions. In particular, our theory has allowed us to determine solute trapping across the lifetime of a filter whose pores are constricting due to contaminant removal. We have also solved the inverse problem of calculating the initial porosity distributions that lead to a filter that blocks uniformly for given operating conditions. 

To track the evolution of filter porosity, we have accounted for a macroscale variation in our filter, using the generalized homogenization technique of \citet{Bruna2015diffusion}. The resulting macroscale equations allow us to efficiently quantify filter performance. By performing a parameter sweep over the functional space of initially linear porosity functions, we have quantified the experimental observation that filters with an initially negative porosity gradient are more effective at removing contaminant over time than filters with an initially zero (or, worse, positive) porosity gradient.

Additionally, our asymptotic reductions in the limit of slow filter blocking and weak spatial porosity variation have allowed us to solve an inverse problem to calculate the initial filter porosity distributions that lead to a filter that blocks uniformly for given operating conditions. We show that these filters provide a greater contaminant removal than a linearly graded filter with the same initial average porosity, but there is a maximum initial average porosity that these filters can reach. As a result, we term these pseudo-optimal filters.

In \citet{dalwadi2015understanding} we showed that, for contaminants that did not alter the geometry of the filter as they adhered, a negative porosity gradient could result in a near-uniform contaminant removal in space, compared to a significantly asymmetric removal for a positive porosity gradient. We conjectured that, were this contaminant removal to result in pore constriction over time, an initially negative porosity gradient would outlast an initially positive porosity gradient. In this paper, we have confirmed and quantified this conjecture. Furthermore, we were able to deduce that a filter with an initially negative porosity gradient yields a larger cumulative contaminant adsorption than an initially positive porosity gradient at a given time, even before blocking occurs. 

In this paper we considered a microscale structure consisting of cylindrical fibres, resulting in a method of varying the local porosity through a single parameter (the fibre radius), and an explicit macroscale equation. However, the homogenization procedure we use can be readily extended to a general microstructure, and one could combine this work with that of~\cite{richardson2011derivation}, who considered a general curvilinear coordinate transform to map a near-periodic microscale to a periodic domain, thus allowing a homogenization method to be applied. In general, the coefficients within the resulting governing equations would have to be determined at each point in space, and thus the resulting macroscale equations are more complicated to solve than in the case presented here. However, there is a computationally efficient middle ground between cubes and a general microstructure. For instance, one could choose a one-parameter family for the potential microstructure, which allows the macroscale coefficients to be written as functions of the porosity. 

One simplifying assumption that we have made in this work is that the contaminant trapping rate is linearly dependent on contaminant concentration, and that the trapping will continue until the microscale cell is fully blocked. In many adsorption models, different trapping conditions, for example, the Langmuir adsorption model, are used (see, for example, \citet{morel1993principles}). This model assumes that there are a finite number of adsorption sites on the obstacle surface (thus, the number of sites is proportional to the obstacle surface area), and that the adsorption layer is only one contaminant molecule thick. Thus, while the adsorption rate will be approximately linearly dependent on the contaminant concentration when few adsorption sites are in use, the adsorption rate will decrease to zero as more adsorption sites are used. This adsorption model can easily be included in this work, and the restriction on the number of adsorption sites means that blocking would not occur in most cases. In this case, the filter that maximizes adsorption would be the one with the largest number of adsorption sites, that is, the filter with the maximum surface area.

While we have shown how to maximize contaminant removal across the functional space of filters with the same initial average porosity, our results only apply up to some maximum initial average porosity. This is because filters that block everywhere at once will have an initially decreasing porosity, and the porosity can never exceed one. Thus, the full control problem for \emph{any} given initial average porosity is still an open problem. This is an important question that would be useful to tackle in future work.

Finally, we note that this work has the potential not only to guide filter manufacture and operating conditions, but also to provide assistance to many other industries. We have introduced a general framework in this paper, which applies to any problem where the underlying transport satisfies an advection--diffusion equation with a general adsorption condition on the microstructure surface. For example, our model can predict drug transport and delivery to tumours, and a simple change of sign to the evolution equation for the porosity will allow prediction of the resultant tumour shrinkage. Using appropriate parameter values, one can predict the effect of various drugs, significantly aiding the task of testing new drug therapies.

\section*{Acknowledgements}

This work was carried out whilst MPD was funded by an EPSRC Impact Acceleration Account Award (grant number EP/K503769/1) at the University of Oxford. IMG is supported by a Royal Society Fellowship. MB is funded by a Junior Research Fellowship of St Johns College, Oxford. This work was also supported by the 2020 Science programme, which was funded through the EPSRC Cross-Disciplinary Interface Programme (grant number EP/I017909/1).

\appendix

  \renewcommand{\theequation}{A\arabic{equation}}
  \setcounter{equation}{0}  

\section{Asymptotic results for the concentration field}

\label{App: asymptotic coefficients}

The functions $\creal_0$ and $\creal_1$ are defined as
\begin{subequations}
\begin{align}
\creal_0(x,t) &= 2 \beta(t) \pav(t) \exp(a(t) x)\left[b(t) \cosh a(t)b(t)(x-1) - \sinh a(t)b(t)(x-1)\right], \\
\creal_1(x,t) &= \dfrac{\exp(a(t)x)}{\sinh a(t)b(t)}\bigg[\left(A_1(x,t) + \beta(t) B_1(t)\right) \sinh \left(a(t)b(t)x\right) \notag \\
&\qquad + \left(A_2(x,t) + \beta(t) B_2(t)\right) \sinh \left(a(t)b(t)\left(x -1\right)\right)\bigg],
\end{align}
\end{subequations}
where
\begin{subequations}
\begin{align}
a(t) &= \dfrac{U(t)}{2 \pav(t) \Deff(\pav)}, \\
b(t) &= \sqrt{1 + f(\pav) /\left(a^2(t) \Deff(\pav) \right)}, \\
\beta(t) &= \dfrac{1}{(1+b^2(t))\sinh a(t) b(t) + 2 b(t) \cosh a(t) b(t)}, \\
A_1(x,t) &= \dfrac{1}{a(t) b(t) \Deff(\pav)}\int_x^1 \! g(\xi,t,\creal_0(\xi,t)) \exp(-a\xi) \sinh ab(\xi-1)\, \mathrm{d}\xi , \\
A_2(x,t) &= \dfrac{1}{a(t) b(t) \Deff(\pav)}\int_0^x \! g(\xi,t,\creal_0(\xi,t)) \exp(-a\xi) \sinh ab\xi \, \mathrm{d}\xi, \\
\begin{pmatrix}
  B_1(t) \\
  B_2(t)
 \end{pmatrix}
 &=
 \begin{pmatrix}
  -q(t) & b(t) \\
  b(t) & -q(t)
 \end{pmatrix}
 \begin{pmatrix}
  b(t) \left(A_2(1,t) - 2 \pbyp{\pc(1,t)}{x}\right) \\
  b(t) A_1(0,t) + N(0,\creal_0(0,t))/(a(t) \Deff(\pav))
 \end{pmatrix}, \\
 g(x,t,\creal) &= -\pbyp{}{x}N(x,t,\creal) + \pc f(\pav) \creal, \\
 N(x,t,\creal) &= \pc \Deff(\pav)\pbyp{\creal}{x} - \dfrac{\creal}{\pav}\left(\Deff(\pav) \pbyp{\pc}{x} - \dfrac{\pc}{\pav} \right), \\
 q(t) &= \sinh \left(a(t)b(t)\right) + b(t) \cosh  \left(a(t)b(t)\right).
\end{align}
\end{subequations}

  \renewcommand{\theequation}{B\arabic{equation}}
  \setcounter{equation}{0}  

\section{Uniformity of removal}
\label{sec: Uniformity of removal}

\begin{figure}
\centering
\includegraphics[width=\textwidth]{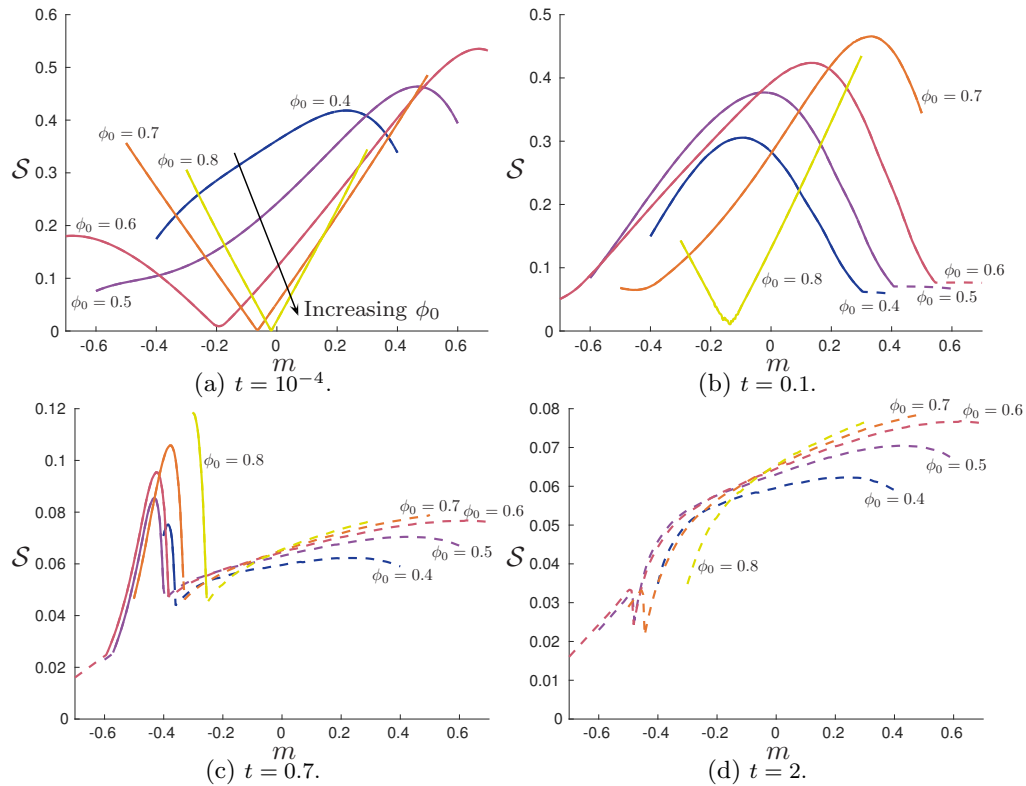}
\caption{(Colour online) The instantaneous uniformity of adsorption, $\mingrad$, defined by \eqref{eq:S}, for varying $\porosity_0$ and $m$ in initial porosity distributions of the form $\porosity(x) = \porosity_0 + m(x - 0.5)$, determined by solving \eqref{eq: 1D flow}--\eqref{eq: additional systems BC}. Each curve represents a different value of $\porosity_0$, which increments in steps of $0.1$ from $\porosity_0 = 0.4$ to $\porosity_0 = 0.8$ and the arrow in (a) denotes curves of increasing $\porosity_0$. We vary $m$ for a given $\porosity_0$ such that $\porosity \in [0.2,0.95]$. Therefore, the available range of $m$ varies with $\porosity_0$. We use the reference values $\porosity = 0.6$, $\Pec = 5$, $\kk = 0.1$ from which to modify appropriate parameters. The dashed curves denote that the minimum pore size has reached $0.01$ and the filtration has stopped.}
\label{fig: S comparisons}
\end{figure}

In \citet{dalwadi2015understanding}, blocking was not explicitly accounted for, and thus the superior performance of filters with a negative porosity gradient was accounted for by introducing a metric that measured the uniformity of uptake. The equivalent metric in our time-dependent case is
\begin{align}
\label{eq:S}
\mingrad(t) = \dfrac{1}{\kk}\int_0^1 \! \left| f(\porosity(x,t)) \creal(x,t) - \int_0^1 \! f(\porosity(s,t)) \creal(s,t) \, \mathrm{d}s \right| \, \mathrm{d}x \geq 0,
\end{align}
where a lower value of $\mingrad$ represents a more uniform removal and $\mingrad = 0$ represents uniform uptake across the filter. As with the cumulative removal, for short times we observe similar results for $\mingrad$ as obtained in \citet{dalwadi2015understanding} (figure~\ref{fig: S comparisons}a), with initial negative porosity gradients generally yielding more uniform removal than positive porosity gradients. However, as time increases, this measure becomes much less useful (figure~\ref{fig: S comparisons}b--d). Whilst lower values of $\mingrad(t)$ are useful indicators of larger removal, they do not necessarily correspond to the largest values of removal (comparing figure~\ref{fig: S comparisons}c to figure~\ref{fig: T comparisons}c). The instantaneous nature of $\mingrad$ does not convey enough global information about the cumulative removal, and factors such as a lower average porosity are also important.

\bibliographystyle{plainnat}
\small
\bibliography{reference}

\end{document}